\documentclass[aps,twocolumn,floatfix,superscriptaddress]{revtex4-2}
\usepackage{graphicx}
\usepackage{dcolumn}
\usepackage{bm}
\usepackage{amssymb}
\usepackage{amsmath}
\usepackage{siunitx}
\usepackage{subfigure}
\usepackage{physics}
\usepackage{float}
\usepackage{multirow,tabularx}
\usepackage{sublabel}
\usepackage[utf8]{inputenc}
\usepackage{hyperref}
\usepackage{mathtools}
\usepackage{float}
\begin{document}

%\preprint{APS/123-QED}

\title{Silicene for flexible electronics}
\author{Swastik Sahoo}
\altaffiliation{These authors contributed equally}
\affiliation{Department of Electrical Engineering, Indian Institute of Technology Bombay, Powai, Mumbai-400076, India}

\author{Abhinaba Sinha}
\altaffiliation{These authors contributed equally}
\affiliation{Department of Electrical Engineering, Indian Institute of Technology Bombay, Powai, Mumbai-400076, India}

\author{Namitha Anna Koshi}
\affiliation{Indo-Korea Science and Technology Center, Bengaluru-560064, India}

\author{Seung-Cheol Lee}
\affiliation{Indo-Korea Science and Technology Center, Bengaluru-560064, India}

\author{Satadeep Bhattacharjee}
\affiliation{Indo-Korea Science and Technology Center, Bengaluru-560064, India}

\author{Bhaskaran Muralidharan}
\email[E-mail:~]{bm@ee.iitb.ac.in}
\affiliation{Department of Electrical Engineering, Indian Institute of Technology Bombay, Powai, Mumbai-400076, India}

\begin{abstract}
The outstanding properties of graphene have laid the foundation for exploring graphene-like two-dimensional systems, commonly referred to as 2D-Xenes.
Amongst them, silicene is a front-runner owing to its compatibility with current silicon fabrication technologies. Recent works on silicene have unveiled its useful electronic and mechanical properties. The rapid miniaturization of silicon devices and the useful electro-mechanical properties of silicene necessitates the exploration for potential applications of silicene flexible electronics in the nano electro-mechanical systems. Using a theoretical model derived from the integration of \textit{ab-initio} density-functional theory and quantum transport theory, we investigate the piezoresistance effect of silicene in the nanoscale regime.
Like graphene, we obtain a small value of piezoresistance gauge factor of silicene, which is sinusoidally dependent on the transport angle. The small gauge factor of silicene is attributed to its robust Dirac cone and strain-independent valley degeneracy. Based on the obtained results, we propose to use silicene as an interconnect in flexible electronic devices and a reference piezoresistor in strain sensors. This work will hence pave the way for exploring flexible electronics applications in other 2D-Xene materials.
\end{abstract}
\maketitle
\section{Introduction}
The unique electrical~\cite{novoselov2004electric,neto2009electronic}, mechanical~\cite{lee2008measurement,liu2007ab,kim2009large} and chemical properties~\cite{geim2009graphene,rosas2011first} of graphene have paved the way for exploration of graphene analogues like silicene, germanene, phosphorene, to name a few. These 2D materials have outstanding properties; however, they face a huge challenge in terms of their large-scale growth and compatibility with the current device-fabrication technology. Amongst them, silicene has the advantage over other contenders because of its compatibility with the already established silicon fabrication technology~\cite{tao2015silicene}. Apart from the commercial advantages, silicene also possesses useful electro-mechanical properties ~\cite{houssa2010electronic,roman2014mechanical} along with additional features such as  strong spin-orbit coupling~\cite{liu2011quantum}, giant magneto-resistance~\cite{rachel2014giant}, and tunable bandgap due to applied electric field~\cite{ni2012tunable}.\\
\indent In the last few years, graphene has been rigorously explored for straintronics applications due to its excellent electro-mechanical properties~\cite{kim2009large,farjam2009comment,ni2008uniaxial,ribeiro2009strained,choi2010effects,klimov2012electromechanical,khan2017mechanical} and exotic physical phenomena resulting from strain maneuvering~\cite{novoselov2004electric,guinea2008midgap,sinha2019piezoresistance,sinha2020graphene,sinha2022ballistic} such as zero-field quantum Hall effect~\cite{guinea2008midgap,levy2010strain}, superconductivity~\cite{si2013first}, and unique Dirac cone dynamics~\cite{farjam2009comment,ni2008uniaxial,ribeiro2009strained}. In many ways, silicene is identical to graphene. Like graphene, silicene exhibits linear energy dispersion relation near the Dirac points \cite{guzman2007electronic, cahangirov2009two}, zero band-gap~\cite{takeda1994theoretical}, and more importantly, a dynamically stable material~\cite{cahangirov2009two}. \\
\indent Despite these advantages, straintronics application of silicene remains unexplored. In this paper, we explore the straintronics properties of silicene in the quasi-ballistic transport regime (which corresponds to a length-scale of around $100$~nm - $200$~nm)~\cite{abidin2017effects} for potential applications in future NEMS systems and flexible electronic devices.Various $2$D materials with enhanced electrical, optical and mechanical properties like graphene, display novel applications in NEMS systems such as reference piezoresistor ~\cite{sinha2020graphene} and ultra high-pressure sensitivity~\cite{sinha2022ballistic,smith2013electromechanical,wagner2018highly}. Because of the similarities between silicene and graphene, the former is expected to contribute to the field of NEMS sensors. The applications of flexible electronics are pretty diverse, with fundamental criteria being robust electronic and excellent mechanical response to strain~\cite{harris2016flexible} followed by portability and manufacturability ~\cite{wong2009materials}, which are expected in silicene~\cite{tao2015silicene,houssa2010electronic,roman2014mechanical} \\
\indent Using theoretical models derived from the integration of \textit{ab-initio} density-functional theory and quantum transport theory, we investigate the piezoresistance effect of silicene in the nanoscale regime. Like graphene, we obtain a small value of piezoresistance gauge factor of silicene, which is sinusoidally dependent on the transport angle. The small gauge factor of silicene is attributed to its robust Dirac cone and strain-independent valley degeneracy. Based on the obtained results, we propose silicene as an interconnect in flexitronic devices and a reference piezoresistor in strain sensors. In the subsequent sections, we describe the mathematical models to extract the hopping parameters of strained silicene and calculate the gauge factors (GFs) along different transport directions.
\section{Simulation Setup} \label{section_2}
\begin{figure}[t]	
	\subfigure[]{\includegraphics[height=0.20\textwidth,width=0.200\textwidth]{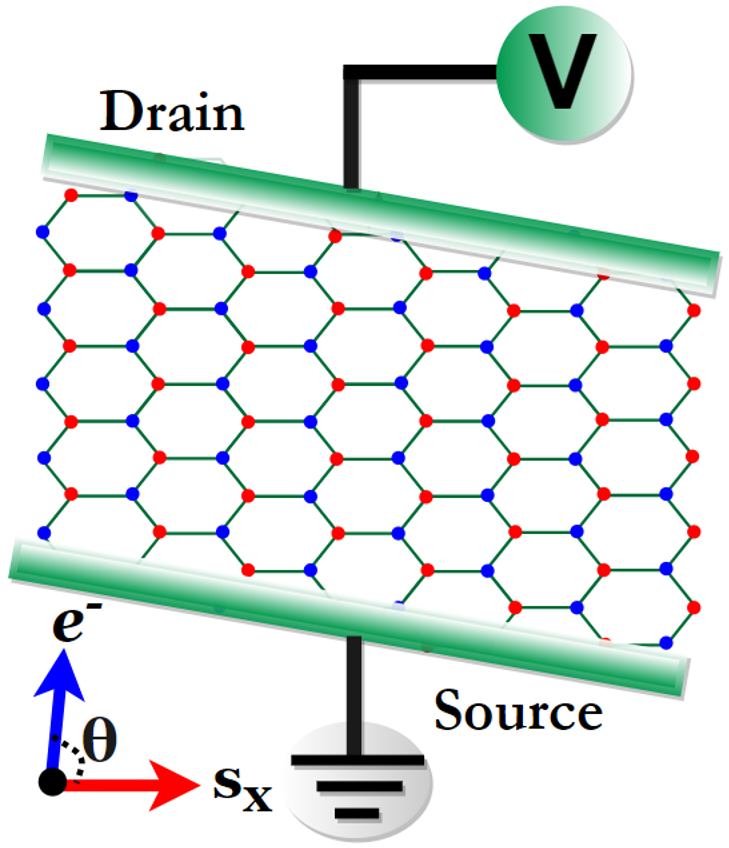}\label{P01_1a}}
	\quad
	\subfigure[]{\includegraphics[height=0.20\textwidth,width=0.218\textwidth]{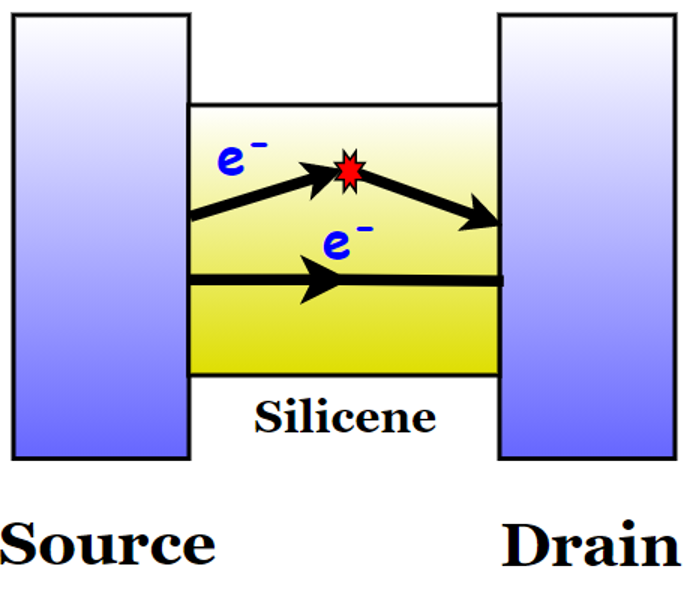}\label{P01_1b}}
	\quad
\caption{Schematic diagrams (a) depicting the piezoresistance setup and (b) the quasi-ballistic transport of electrons in silicene.}
\label{P01_1}
\end{figure}

The device setup for the calculation of piezoresistance gauge factor of silicene along different transport angles is shown in Fig.~\ref{P01_1a}. It consists of a silicene sheet placed between the two contacts, source and drain. A uniaxial strain of magnitude ($s$) varying from $0\%$ to $5\%$ is applied along the armchair (AC) and zigzag (ZZ) directions, and the resistance is simultaneously measured along different transport angles `$\theta$' ($0$$^{\circ}$ to $90$$^{\circ}$) for applied voltage lying between $\pm 10$ meV.\\
\indent The device dimension along the transport direction is nearly $100$~nm. Hence, the electrons undergo quasi-ballistic transport.
Figure~\ref{P01_1b} shows quasi-ballistic transport of electrons between the source and drain. In the subsequent sub-sections, we obtain the tight-binding parameters of the band structure of strained silicene using density functional theory and further obtain the gauge factor using Landauer quantum transport formalism.

\subsection{Extraction of hopping parameters of silicene}
\begin{figure}[!t]	
	\subfigure[]{\includegraphics[height=0.20\textwidth,width=0.255\textwidth]{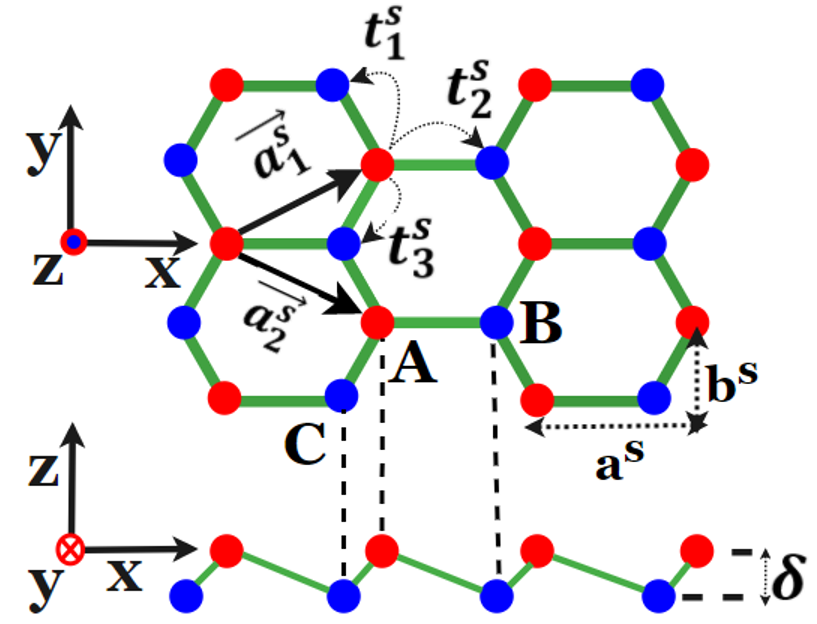}\label{P01_2a}}
	\quad
	\subfigure[]{\includegraphics[height=0.20\textwidth,width=0.200\textwidth]{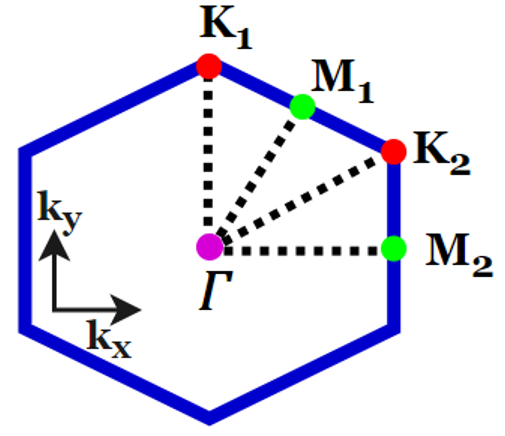}\label{P01_2b}}
	\quad
\caption{(a) Top view and side view of silicene crystal lattice with $\protect\overrightarrow{a_1^{s}}$  and $\protect\overrightarrow{a_2^{s}}$ as the primitive vectors, and $\delta$ as the buckling constant.  A and B represent the non co-planar sub-lattices.  ${t_1^{s}}$, ${t_2^{s}}$ and ${t_3^{s}}$ represent the nearest neighbour tight binding parameters.  (b) Schematic diagram depicting the high symmetry path ${\mathrm{M_{1} \Gamma M_{2} K_{2} M_1 K_{1}\Gamma K_{2}}}$ in the $1^{st}$ Brillouin zone of strained silicene}
\label{P01_2}
\end{figure}

\begin{figure}[t]	
	\subfigure[]{\includegraphics[height=0.18\textwidth,width=0.229\textwidth]{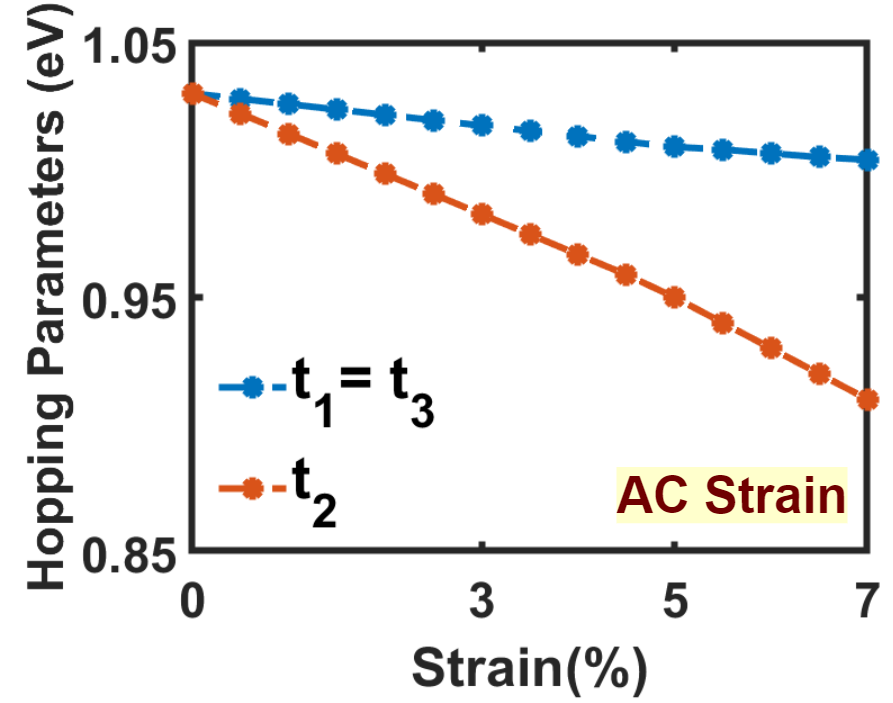}\label{P01_3a}}
	\quad
	\subfigure[]{\includegraphics[height=0.18\textwidth,width=0.229\textwidth]{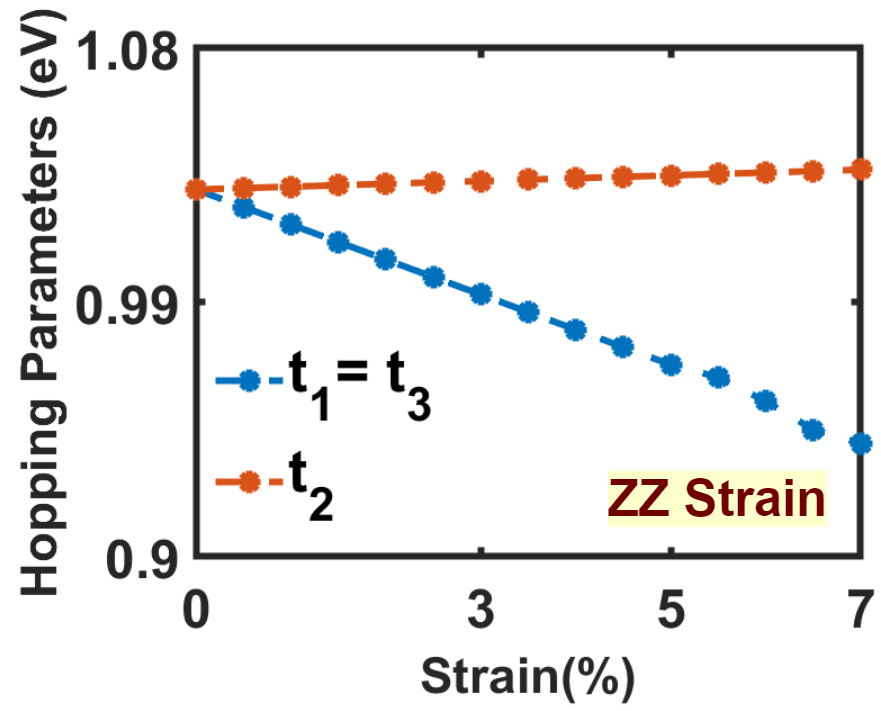}\label{P01_3b}}
	\quad
\caption{Variation of the nearest neighbour tight binding parameters as a function of strain along the (a) armchair, and (b) zigzag directions.}
\label{P01_3}
\end{figure}

Silicene has a buckled honey-comb structure~\cite{takeda1994theoretical}. It consists of two triangular sub-lattices, shown in Fig.~\ref{P01_2a} by red (denoted by A) and blue (denoted by B) dots respectively. These sub-lattices are non co-planar; which gives silicene a buckled honey-comb structure. Figure~\ref{P01_2b} depicts the first-Brillouin zone of silicene. For a uniaxially strained silicene, the high symmetry path is given by $\mathrm{M_{1}\Gamma M_{2} K_{2} M_1 K_{1}\Gamma K_{2}}$. Using $\textit{ab-initio}$ calculations, we obtain the band structure of strained silicene along this path in the linear elastic regime~\cite{peng2013mechanical}. 
 For $10$ values of strain ($\leq$ 7\%), we carry out band structure calculations and obtain the nearest neighbour tight binding (TB) parameters as explained below. The appropriate TB parameters for other (intermediate) values of strain are taken from the line obtained by fitting. The Slater-Koster (SK) parameter, $V_{pp\pi}$ is computed by fitting DFT data to a tight binding (TB) model as implemented in by Nakhaee \textit{et al.}~\cite{nakhaee2020tight}. We model the SK parameter of strained and unstrained silicene by the following exponential relation. \\
\begin{equation} 
    V^{'}_{pp\pi} = V^{0}_{pp\pi} e^{-\beta\big(\frac{{\left|\delta\right|}}{a_0} - 1\big)}
    \label{P01_eq1}
\end{equation}
Here $\beta$ is a parameter that determines the effect of strain on the SK parameter. For the $\pi$ band of graphene, it is around 3 \cite{masir2013pseudo,pereira2009tight}. The numerical value of $\beta$ for the SK parameter, $V_{pp\pi}$ is estimated by fitting the band structure in the linear dispersion region $\pm$0.2 eV energy window to the TB model. Using this value of $\beta$, we determine the hopping parameter of the strained lattice using the relation. \\
\begin{equation} 
t_n = t_0e^{-\beta\big(\frac{{\left|\delta^\prime_n\right|}}{a_0}-1\big)},
\label{P01_eq2}
\end{equation}
where, $t_{0}$ is the magnitude of hopping amplitude of unstrained silicene and is taken to be 1.03 eV and $\delta^\prime_n$ is the nearest neighbour position vector in the strained lattice. As reported in \cite{chuan2020two}, hopping amplitude is obtained by fitting the parameter to experimental or computational results, we take the current value by bench-marking with our DFT results. In previous studies, the values of `$t_{0}$' used are -1.60 eV~\cite{liu2011low}, -1.02 eV~\cite{chuan2020electronic} and -1.03 eV~\cite{roome2014beyond}. The magnitude of hopping parameters of silicene as a function of strain in the AC and ZZ directions are given in Figs.~\ref{P01_3a} and~\ref{P01_3b} respectively. For uniaxial strains, $t_1 = t_3 \neq t_2$. In the AC direction, $t_1, t_2$ and $t_3$ decreases whereas for strain in the ZZ direction, $t_2$ increases and $t_1, t_3$ decreases, with  increase in the magnitude of strain.
The nearest-neighbour tight binding energy dispersion relation for strained silicene is given by, 
\begin{equation} 
E(k)= \pm |t_2+t_1^{s} e^{-(i \vec{k} \cdot \overrightarrow{a_1^{s}})}+t_3^{s} e^{-(i\vec{k} \cdot \overrightarrow{a_2^{s}})}|.
\label{P01_eq3}
\end{equation}
\subsection{Quantum Transport Model}
\begin{figure*}[!t]
\subfigure[]{\includegraphics[height=0.25\textwidth,width=0.25\textwidth]{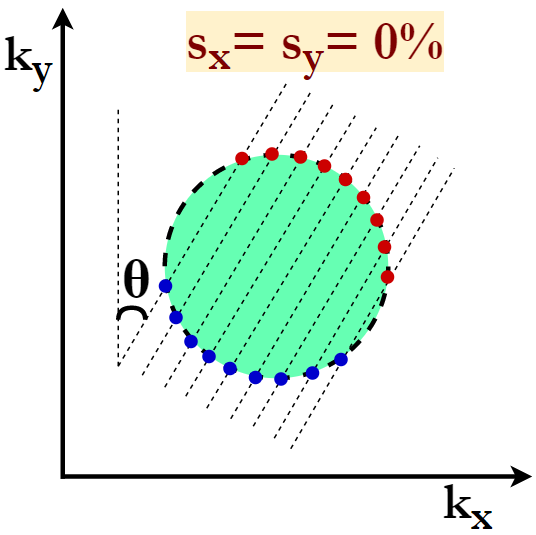}\label{P01_4a}}
\quad
\hspace{0.75 cm}
\subfigure[]{\includegraphics[height=0.25\textwidth,width=0.25\textwidth]{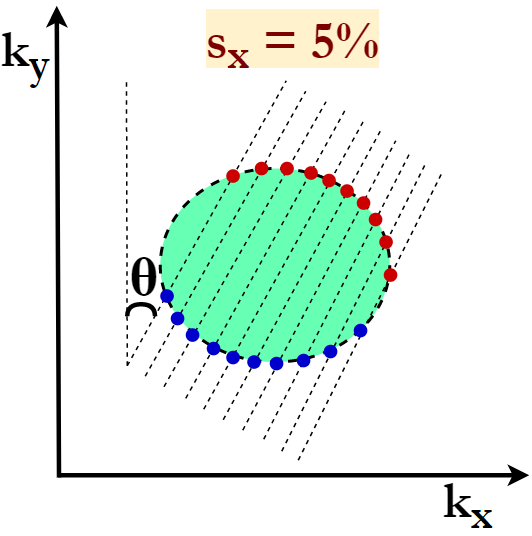}\label{P01_4b}}
\quad
\hspace{0.8 cm}
\subfigure[]{\includegraphics[height=0.25\textwidth,width=0.25\textwidth]{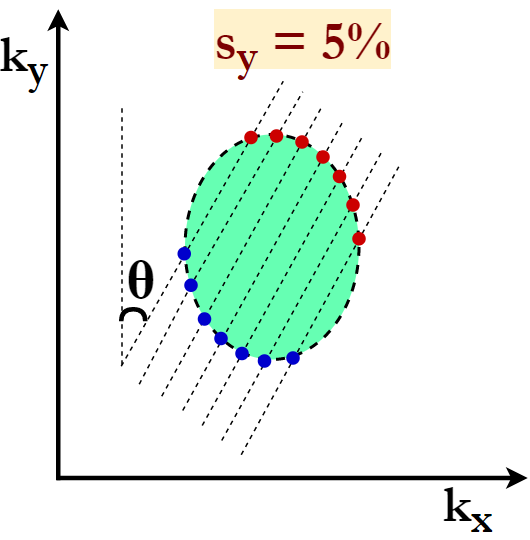}\label{P01_4c}}
\caption{Schematic diagram depicting the modes in constant energy surfaces at (a) 0\% strain, (b) 5\% strain along the armchair direction, and (c) 5\% strain along the zigzag direction. The dotted lines represent the direction of transverse modes, and the red and blue dots represent the modes for forward and backward propagating electrons respectively.}
\label{P01_4}
\end{figure*}

\begin{figure*}[htbp]
\subfigure[]{\includegraphics[height=0.25\textwidth,width=0.25\textwidth]{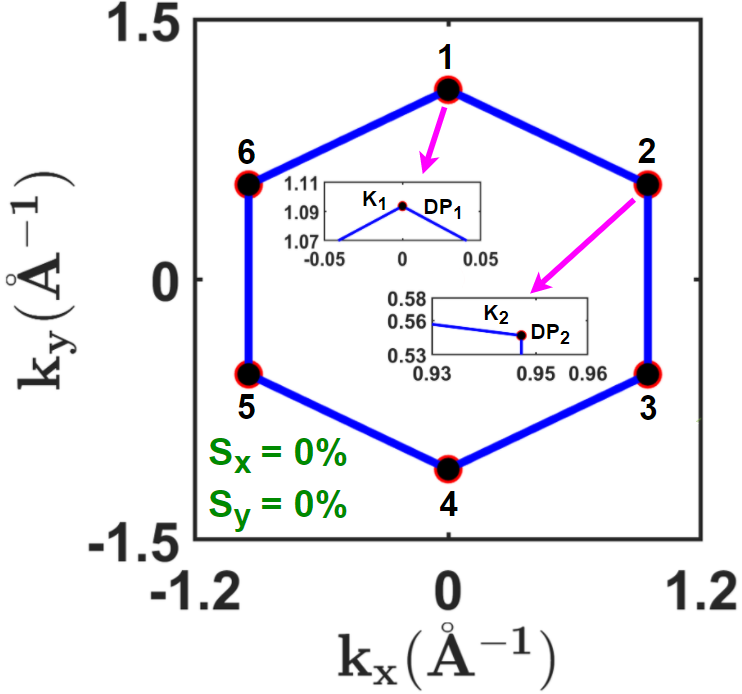}\label{P01_5a}}
\quad
\hspace{0.75 cm}
\subfigure[]{\includegraphics[height=0.25\textwidth,width=0.25\textwidth]{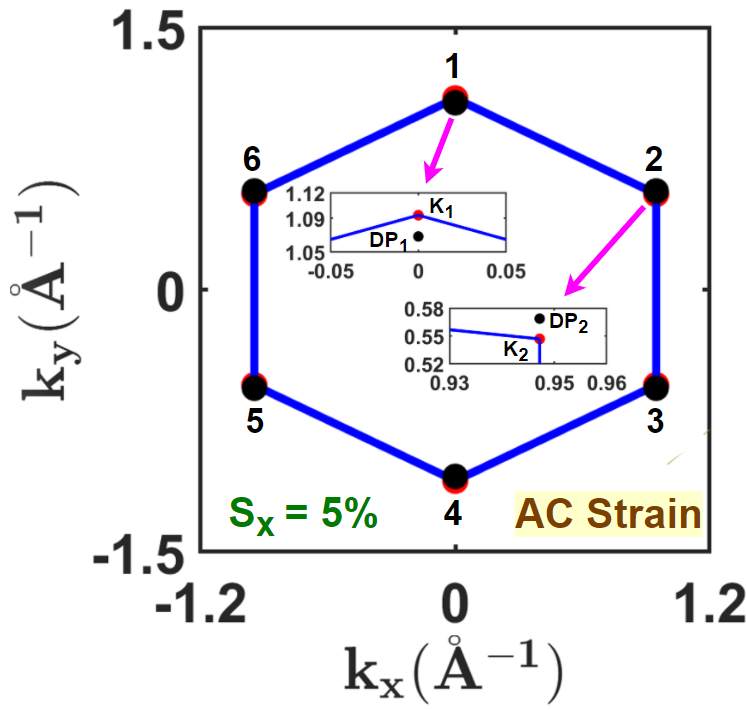}\label{P01_5b}}
\quad
\hspace{0.8 cm}
\subfigure[]{\includegraphics[height=0.25\textwidth,width=0.25\textwidth]{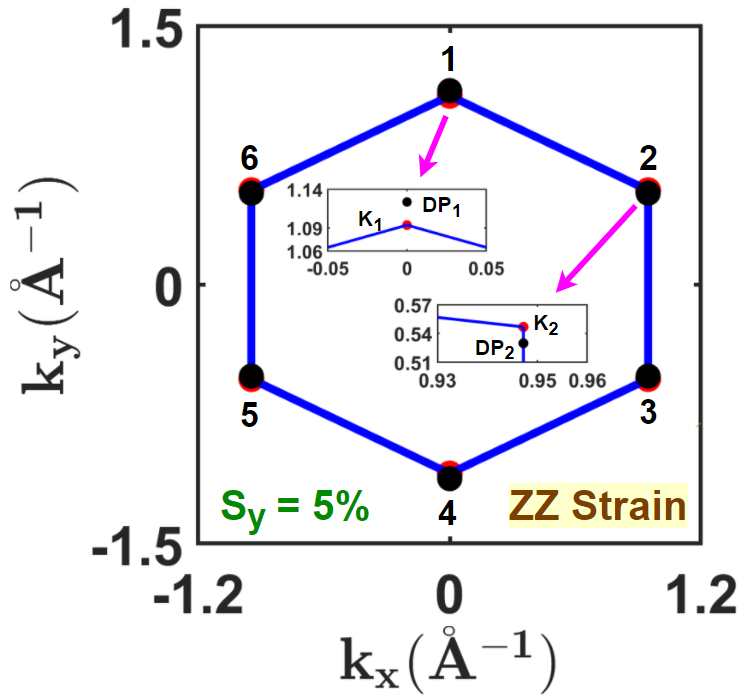}\label{P01_5c}}
\caption{The relative shift in position of Dirac points $\mathrm{DP_{1}}$ and $\mathrm{DP_{2}}$ with respect to $K_1$ and $K_2$ at (a) $0\%$ strain,(b) $5\%$ armchair strain, and (c) $5\%$ zigzag strain along the line joining points 1 and 4, and 2 and 3 respectively. At $0\%$ strain, the Dirac points lie on the K-points, whereas at non-zero strain, the movement of the Dirac points is equal and opposite for armchair and zigzag strains.}
\label{P01_5}
\end{figure*}

Once the TB parameters are obtained, we use the Landauer formula~\cite{datta2018lessons} to calculate current-voltage characteristics of silicene in the quasi-ballistic regime. The Landauer formula is expressed as
\begin{equation} 
 I_\theta^{s} (V)=\frac{2q}{h} \int_{-\infty}^{\infty} {T^s}(E) [f_1(E-\mu_1)-f_2(E-\mu_2)] dE,
 \label{P01_eq4}
\end{equation}

where $f_{1}$ and $f_{2}$ are the Fermi functions, and $\mu_{1}$ and $\mu_{2}$ are the Fermi-levels at the source and drain respectively, $T^{s}$ is the transmission of strained silicene, $q$ is the electronic charge, $h$ is the Planck's constant, $E$ is the energy, and  $I_\theta^{s} (V)$ is the current along the transport direction $(\theta)$ at an applied strain (s). Here, the voltage varies from $-10$~mV to $+10$~mV. \\
\indent The transmission $T^{s}(E)$ is the product of transmission probability $T(E)$ and mode density $M_\theta^{s}(E)$. The mean free path in the linear regime depends on the relaxation time near the Dirac cone~\cite{shao2013first} and in case of silicene it is comparable to the length of the device~\cite{guzman2018transport}. So, the transport in silicene is considered to be in the quasi-ballistic regime. $T(E)$ is expressed as 
\begin{equation} 
T(E)=\frac {\lambda (E)}{\lambda(E)+L_s},
\label{P01_eq5}
\end{equation}
where  $\lambda (E)$ is the mean free path as a function of energy and $L_{s}$ is the length of silicene channel. 

We calculate the mode density of silicene from the band structure using band counting method for 2-D Dirac materials~\cite{sinha2019piezoresistance,sinha2020graphene}. Figure~\ref{P01_4} represents the constant energy surfaces and modes along the transport direction  $(\theta)$ for different strains along AC and ZZ directions. The dotted lines over the constant energy surface represent the transverse modes, and the red and blue dots represent the forward and backward propagating electrons, respectively. 
The Dirac cone degeneracy for silicene in the first Brillouin zone is two~\cite{liu2013dirac,sinha2019piezoresistance}. So, effectively the mode density is calculated by considering the sum of forward and backward moving electrons in a single Dirac cone~\cite{sinha2020graphene}, and is mathematically expressed as
\begin{equation} 
M_\theta^s(E)= 2\times n_\theta^s(E)
\label{P01_eq6}
\end{equation}
where, $n_\theta^s(E)$ is the number of transverse modes passing through the constant energy surfaces at energy `$E$'. Transverse modes have a separation of $2\pi/w_s$, where $w_s$ represents the width of strained silicene sheet and is equal to  $w_s= w(1+\frac{s}{100})$. Here $w$ represents the width of the pristine silicene cell. Thus, transmission is mathematically expressed as 
\begin{equation} 
 T^{s}(E)= T(E)*M_\theta^s(E).
 \label{P01_eq7}
\end{equation}
The contacts are assumed to be ideal in our calculations. The energy range for this Fermi-Dirac distribution is from $-0.2$ eV to $+0.2$ eV, which is well within the linear regime.  The resistance is calculated from the Landauer formalism, and is written as 
\begin{equation} 
R_\theta^{s}=\frac{dV}{dI_\theta^{s} (V)}
\label{P01_eq8}
\end{equation}
The above equation is used to calculate the resistance at different strain and transport angles. Using these values, we can calculate Angular Gauge Factor (AGF). AGF is expressed as: 
\begin{equation} 
(AGF)_\theta^{s}=\frac{(R_\theta^{s}-R_\theta^{0})}{s*R_\theta^{0}},
\label{P01_eq9}
\end{equation}
where, ${R_\theta}^{0}$ is the resistance at zero strain and ${R_\theta}^{s}$ is the resistance at a particular value of strain $s$.
As the strain is applied throughout the transport angle $(\theta)$ so, the AGF is the average of all GFs at different strains along a particular transport angle. Thus, the average AGF is expressed as:  
\begin{equation} 
(AGF)_\theta= \overline {(AGF)_\theta^{s}}
\label{P01_eq10}
\end{equation}

\section{Results and discussion}

In this section, we obtain the piezoresistance gauge factor of silicene along different transport angles using the theoretical models discussed in the previous section and rationalize our findings in terms of the change in transmission due to the shifting and deformation of the Dirac cones. Further, we discuss the implication of our results in flexible electronics devices.

\subsection{Effect of uniaxial strain on Dirac cones}
Here, we analyze the effect of uniaxial strain on the Dirac points of silicene. Energy close to the Dirac points is responsible for electronic transport in the linear regime. In order to understand the piezoresistance effect in silicene, we need to understand how the Dirac points behave in silicene due to strain. \\
\indent From the DFT band structure (see Fig.~\ref{P01_5}), we see a shift in the position of Dirac points from their respective K-points due to strain. A similar shift of the Dirac cones in silicene due to uniaxial strain is predicted by Farokhnezhad~\textit{et al.}~\cite{farokhnezhad2017strain}. The shift in the Dirac points $\mathrm{DP_{1}}$ and $\mathrm{DP_{2}}$ with respect to $\mathrm{K_{1}}$ and $\mathrm{K_{2}}$ at zero strain and 5\% strain along the AC and ZZ directions are shown in Figs.~\ref{P01_5a}, ~\ref{P01_5b} and ~\ref{P01_5c} respectively. The response of the Dirac points $\mathrm{DP_{1}}$ and $\mathrm{DP_{4}}$, $\mathrm{DP_{2}}$ and $\mathrm{DP_{3}}$, and $\mathrm{DP_{5}}$ and $\mathrm{DP_{6}}$ are identical on application of strain along the AC and ZZ directions~\cite{sinha2019piezoresistance}. For strain along the AC direction, Dirac points denoted as 1 and 4 shifts inside the Brillouin zone along the line joining the points $\mathrm{K_{1}-\Gamma}$ (see~Fig.~\ref{P01_5b}) and rest of the Dirac points 2 and 3, and 5 and 6 shifts outside of the Brillouin zone along the line joining the points $\mathrm{K_{2}-K_{3}}$, and $\mathrm{K_{5}-K_{6}}$ respectively (see~Fig.~\ref{P01_5b}).
\begin{figure}[t]
  \centering
   \includegraphics[width=0.48\textwidth]{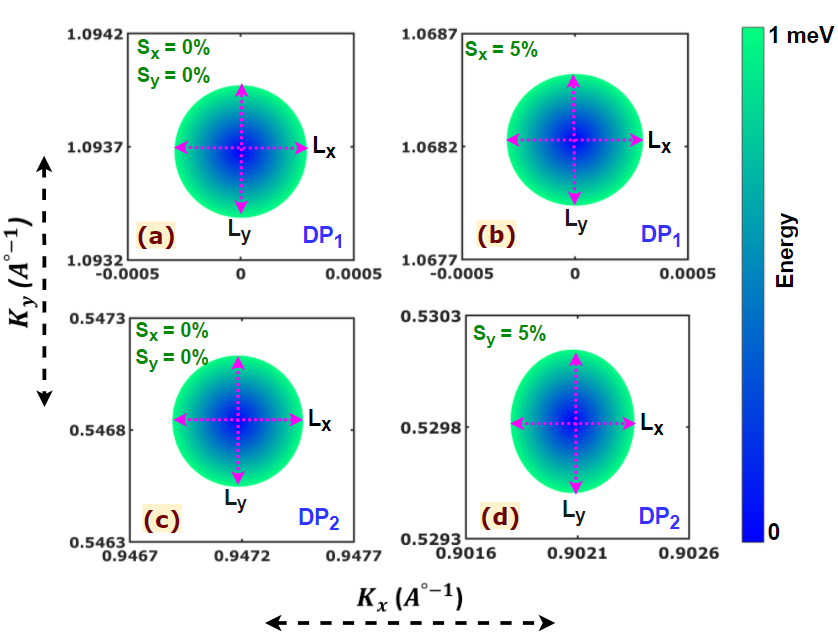}
   \caption{Energy color-map near the Dirac points: $\mathrm{DP_{1}}$ at (a) $0\%$ and (b) $5\%$ armchair strain, and $\mathrm{DP_{2}}$  at (c) $0\%$ and (d) $5\%$ zigzag strain.}
   \label{P01_6}
\end{figure}
 In contrary to AC strain, strain along the ZZ direction results in shifting of Dirac points 1 and 4 shifts away from the Brillouin zone, whereas the other Dirac points 2 and 3, and 5 and 6 shift inside and move closer to each other along the respective edges of the Brillouin zone ( see Fig.~\ref{P01_5c}). The shifting of these two sets of Dirac points is identical and opposite. 
\setlength{\tabcolsep}{4.9pt}
\begin{table}[H]
\caption{Shifting of Dirac cones from the K-points (in \AA$^{-1}$) for armchair and zigzag strain}
    \centering
\begin{tabular}{c cc cc}
 \hline\hline
        &
 \multicolumn{2}{c}{\textit{AC Strain}} & \multicolumn{2}{c}{\textit{ZZ Strain}}\\
 %   \cline{2-5}
        Strain($\%$) & $K_1$ - $\mathrm{DP_{1}}$ &  $K_2$ - $\mathrm{DP_{2}}$ &  $K_1$ - $\mathrm{DP_{1}}$ & $K_2$ - $\mathrm{DP_{2}}$\\[0.05cm]
\hline
 $0$ & $0$ & $0$ & $0$ & $0$\\[0.02cm]
 $2$ & $0.0102$ & $-0.0051$ & $-0.0115$ & $0.0064$\\[0.02cm]
 $5$ & $0.0255$   & $-0.0129$  & $-0.0298$  & $0.0170$ \\[0.05cm]
 \hline\hline
  \end{tabular}
\label{P01_table1}
  \end{table}
Table~\ref{P01_table1} shows the magnitude of the Dirac-point shift from the $\mathrm{K}$-points. Thus, the Dirac points $\mathrm{DP_{1}}$ and $\mathrm{DP_{2}}$ are sufficient to study the effect of strain on the first Brillouin zone, and the overall Dirac cone degeneracy is two irrespective of the magnitude and direction of strain. In addition to shifting of the Dirac points, uniaxial strain deforms the shape of the Dirac cones into oval-shaped cones (see~Fig.~\ref{P01_4}). Unlike graphene, deformation of the Dirac cones are not exactly the same for strain along the AC direction or ZZ direction~(see Fig.~\ref{P01_6})~\cite{sinha2019piezoresistance}.\\
Table~\ref{P01_table2} shows the value of the major and minor axes of the deformed Dirac cones at E= 1 meV.
\setlength{\tabcolsep}{4.5pt}
\begin{table}[H]
  \caption{The major and minor axes of the deformed Dirac cones (in \AA$^{-1}$) at E$=1$ $meV$  for armchair and zigzag strain}
 \centering
\begin{tabular}{c cc cc}
 \hline\hline
        &
 \multicolumn{2}{c}{\textit{AC Strain    }} & \multicolumn{2}{c}{\textit{ZZ Strain}}\\
    %\cline{2-5}
        Strain ($\%$)  & $DP_1 (L_x)$ & $DP_1 (L_y)$ & $DP_2 (L_x)$ & $DP_1 (L_y)$\\[0.1cm]
\hline
 $0$ & $0.0005857$ & $0.0005857$ & $0.0005857$ & $0.0005857$\\[0.05cm]
 $2$ & $0.0005924$ & $0.0005839$ & $0.0005731$ & $0.0006071$\\[0.05cm]
 $5$ & $0.0006035$ & $0.0005821$  & $0.0005545$  & $0.0006431$\\[0.1cm]
 \hline\hline
  \end{tabular}
\label{P01_table2}
  \end{table}
\subsection{Directional piezoresistance calculation}
\begin{figure}[t]
    \subfigure[]{\includegraphics[height=0.18\textwidth,width=0.225\textwidth]{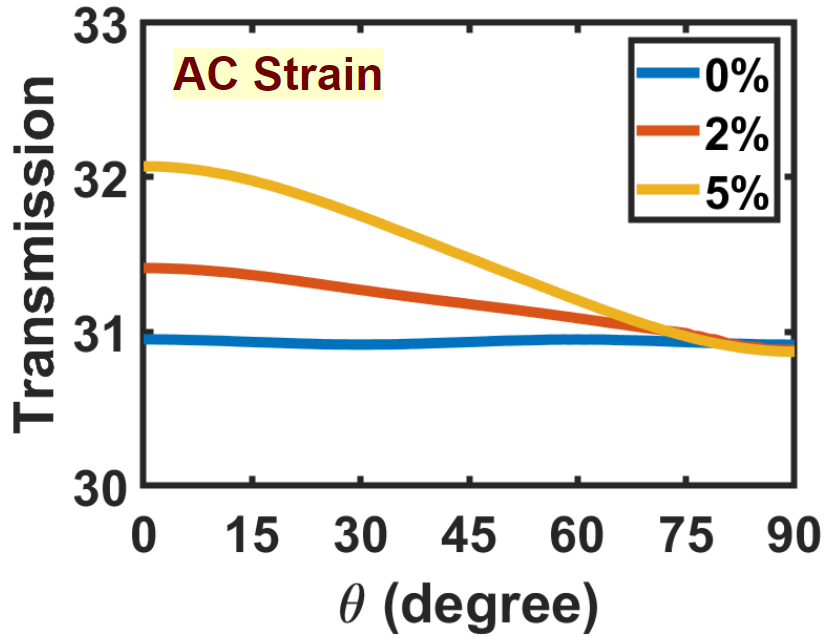}\label{P01_7a}}
\quad
    \subfigure[]{\includegraphics[height=0.18\textwidth,width=0.225\textwidth]{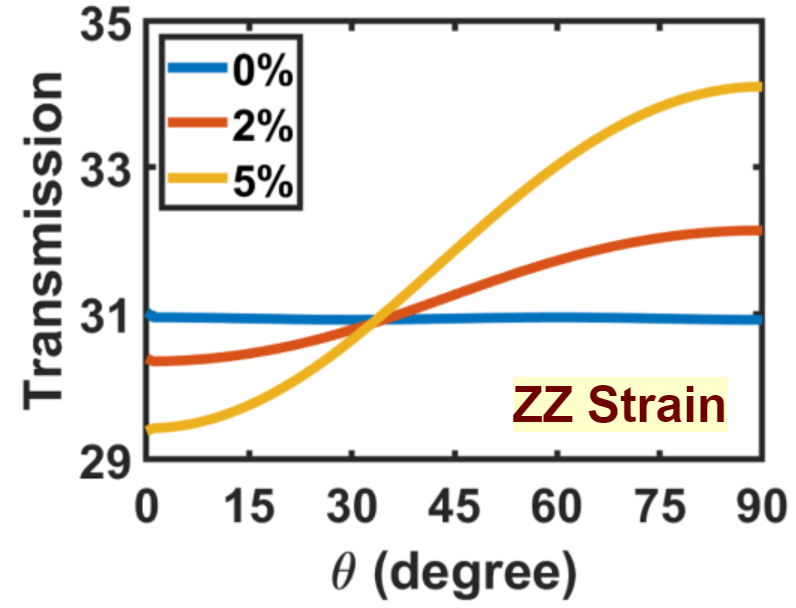}\label{P01_7b}}
\quad
    \subfigure[]{\includegraphics[height=0.18\textwidth,width=0.225\textwidth]{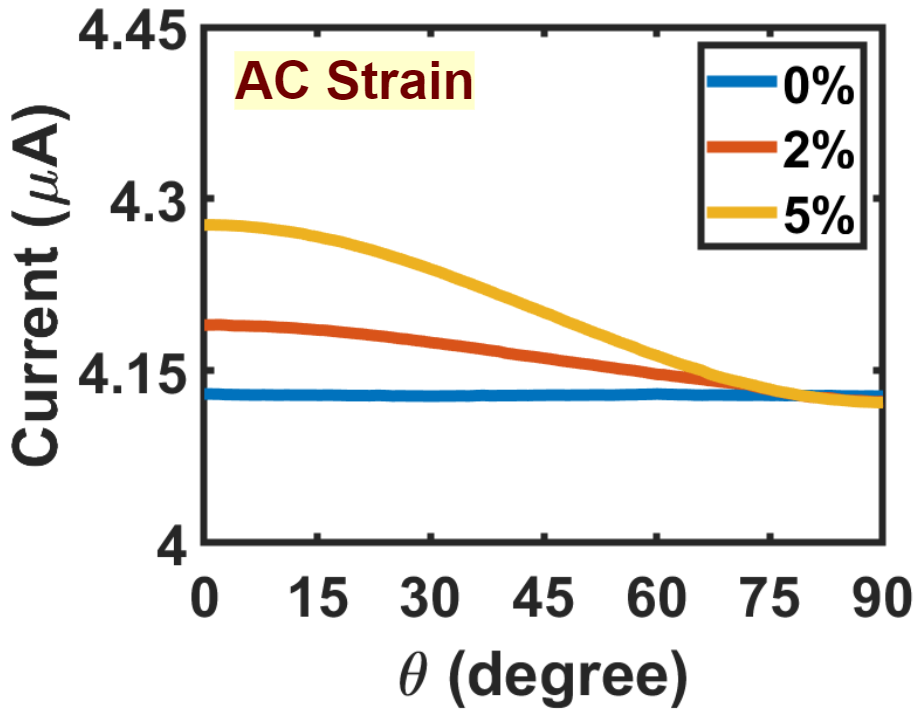}\label{P01_7c}}
\quad
    \subfigure[]{\includegraphics[height=0.18\textwidth,width=0.230\textwidth]{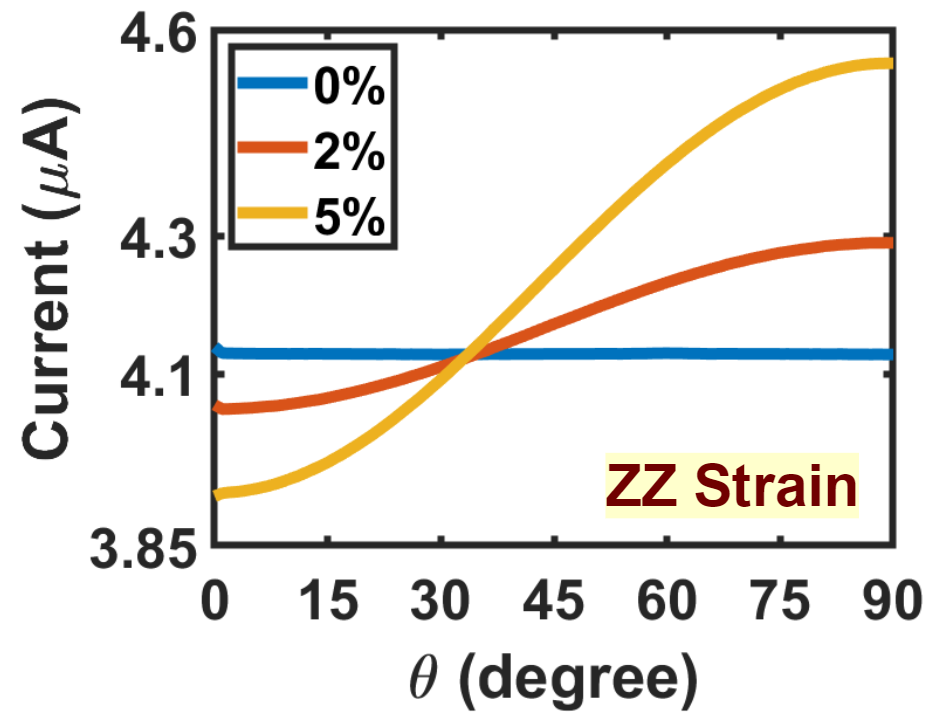}\label{P01_7d}}
\caption{Plots of transmission and current as a function of  transport angle $(\theta)$  varying from $0^{\circ}$  to $90^{\circ}$ at $0\%$, $2\%$ and $5\%$ strain along the armchair and zigzag directions}
\label{P01_7}
\end{figure}

In this sub-section, we obtain the various transport parameters such as transmission, current, resistance and gauge factor as a function of the transport-angle and applied strain along the AC and ZZ directions in the quasi-ballistic transport regime~(since the mean free path of silicene is around $20$~nm~\cite{abidin2017effects}). \\
\indent The transmission of silicene as a function of transport angle $(\theta)$ with $0\%$, $2\%$ and $5\%$ AC strain is shown in Fig.~\ref{P01_7a}. Here, $\theta$ varies from $0$$^{\circ}$ to $90$$^{\circ}$. From the figure, we infer that with an increase in $\theta$, the transmission reduces due to the reduction in the value of mode density. In the case of ZZ strain, the transmission increases with an increase in $\theta$ (see Fig.~\ref{P01_7b}). The plots of current as a function of transport angle for AC and ZZ strains have a similar pattern like the transmission (see Figs.~\ref{P01_7c} and~\ref{P01_7d}). This is because the current is proportional to the transmission at a particular energy [see Eq.~\eqref{P01_eq4}]. 
The transmission remains constant at different transport angles due to circular constant energy surfaces. Due to AC strain, the value of $L_{x}$ increases while $L_{y}$ decreases, as a result, the transmission has a higher value at $\theta=$ 0$^{\circ}$ than unstrained silicene, and it decreases with an increase in the transport angle. On the contrary, the value of $L_{x}$ decreases while $L_{y}$ increases for ZZ strain. Consequently, the transmission along $\theta=$ $0^{\circ}$ has a lower value than the transmission of unstrained silicene along $\theta=$ $0^{\circ}$. Thus, the transmission increases with an increase in the transport angle.\\
\begin{figure}[t]	
	\subfigure[]{\includegraphics[height=0.18\textwidth,width=0.229\textwidth]{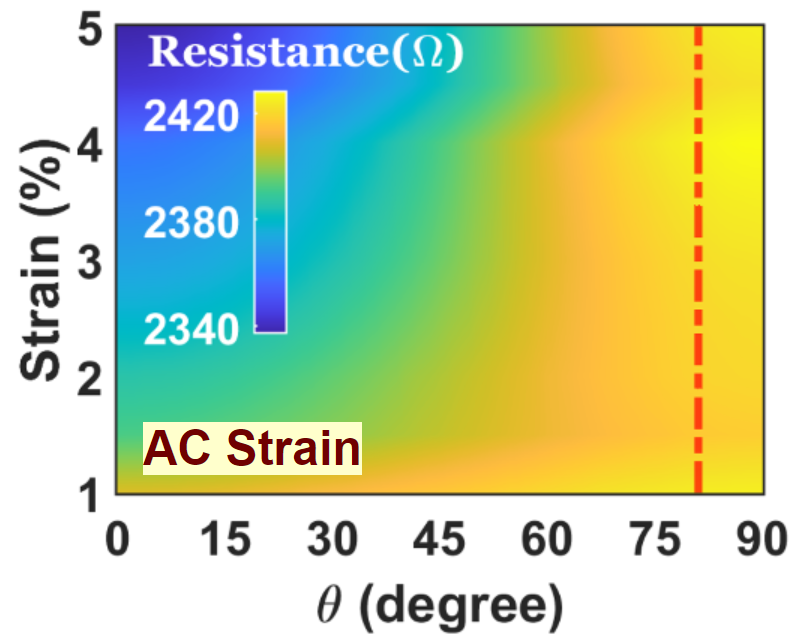}\label{P01_8a}}
	\quad
	\subfigure[]{\includegraphics[height=0.18\textwidth,width=0.229\textwidth]{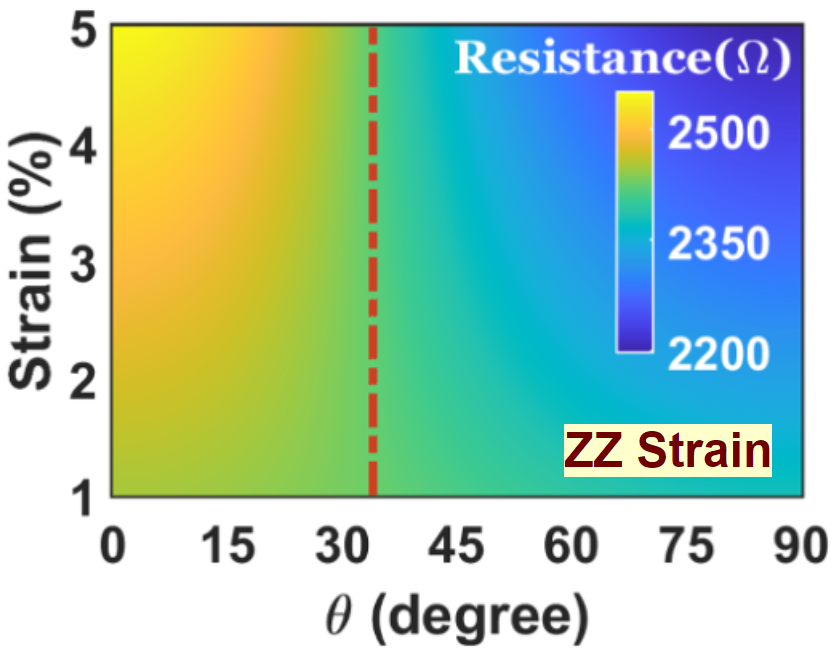}\label{P01_8b}}
	\quad
\caption{Resistance as a function of transport angle $(\theta)$ and percentage strain along the (a) armchair, and (b) zigzag directions. The strain insensitive transport angles are $81^{\circ}$ and $34^{\circ}$  in (a) and (b) respectively}
\label{P01_8}
\end{figure}
Further, we calculate the resistance as a function of strain and transport angle from the I-V characteristics obtained using the Landauer formula [Eq.~\eqref{P01_eq4}]. The plots of resistance versus $\theta$ for AC strain and ZZ strain are shown in Figs.~\ref{P01_8a} and~\ref{P01_8b} respectively. The value of resistivity obtained in this work for a $1~\mu m$ wide silicene in the quasi-ballistic regime is \SI{2.42}{\kilo\ohm} (refer Table~\ref{P01_table3}). Assuming transmission to be equal to one, we obtain resistivity \SI{0.404}{\kilo\ohm}. The value of resistivity of silicene (assuming transmission as 1) is less compared to the resistivity of graphene~\cite{sinha2019piezoresistance}. This is because the number of TMs in silicene is more than graphene (for the sheet-width) due to the larger size of its Dirac cones. 
\begin{table*}
\caption{Comparison of various parameters of silicene and graphene}
\begin{ruledtabular}
\begin{tabular}{cccc}
 &\multicolumn{2}{c}{Silicene}&{Graphene}\\
 Parameters&$T(E)=1$&T(E) $<$ 1&$T(E)=1$
\\ [0.08cm]
\hline
 Dirac cone radius (1~meV) (\AA$^{-1}$)  & $0.00029$ & $0.00029$ & $0.00018$~\cite{sinha2019piezoresistance} \\[0.03cm]
 Current ($\mu$A) &  $24.76$    & $4.12$    &  $15.2$~\cite{sinha2019piezoresistance}\\ [0.03cm]
 Resistance (K$\Omega$) & $0.404$   & $2.42$  & $0.659$~\cite{sinha2019piezoresistance}\\
\end{tabular}
\end{ruledtabular}
\label{P01_table3}
\end{table*}
In Figs.~\ref{P01_8a} and~\ref{P01_8b}, we see that the resistance remain constant at the transport angles $81^{\circ}$ and $34^{\circ}$ for AC and ZZ strains respectively (shown by red dotted lines). This is due to no change in transmission value at the critical angles with strain. Figures~\ref{P01_9a} and~\ref{P01_9b} depict the AGF as a function of the transport angle. The AGFs vary sinusoidally with transport angle for the AC and ZZ strains. The variation of AGF resembles a sinusoidal function similar to the one obtained by Sinha~\textit{et al.} ~\cite{sinha2020graphene} for graphene. The AGF is of the form of
\begin{equation}\label{P01_eqn11}
\mathrm{AGF=-{\{ P\cos(2\theta)+Q}\}}
\end{equation}
where P and Q are constants, and their values are respectively 0.249 and 0.237 for AC strain, and 1.079 and -0.239 for ZZ strain. \\
\subsection{Implications of the results}
Table~\ref{P01_table4} compares the piezoresistance GF of silicene with some other prominent materials. From the table, we infer that the GF of silicene is very small, whereas its 3D counterpart silicon has a very high GF. 
\setlength{\tabcolsep}{16.5pt}
\begin{table}[H]
\caption{Comparisons of the GFs of various materials}
\centering
\begin{tabular}{c c c}\\ 
 \hline\hline
 Material & GF & Reference\\[0.08cm]
 \hline
 Silicene & 0.758 & This work\\[0.03cm]
 Graphene & 0.6 & \cite{sinha2020graphene}\\[0.03cm]
 Silicon & 200 & \cite{smith1954piezoresistance} \\[0.03cm]
 Germanium  & 150 &  \cite{smith1954piezoresistance} \\[0.08cm]
 \hline\hline
\end{tabular}
\label{P01_table4}
\end{table}
 This is due to their different band structures at zero strain and the robustness of the Dirac cones in the presence of strain. The GF of silicene is considerably lower than other semiconductors also. Hence, silicene is not a good choice for piezoresistance strain sensors.
Nevertheless, silicene is an atomically-thin membrane and is expected to have a high value of adhesivity (like most of the similar 2D materials)~\cite{megra2019adhesion} and elasticity~\cite{peng2013mechanical}. Thus, silicene is expected to have a very high-pressure sensitivity like graphene~\cite{sinha2022ballistic} and $\mathrm{PtSe_{2}}$~\cite{Smith2013, Wagner2018} despite a low GF. \\
\begin{figure}[t]
    \subfigure[]{\includegraphics[height=0.18\textwidth,width=0.225\textwidth]{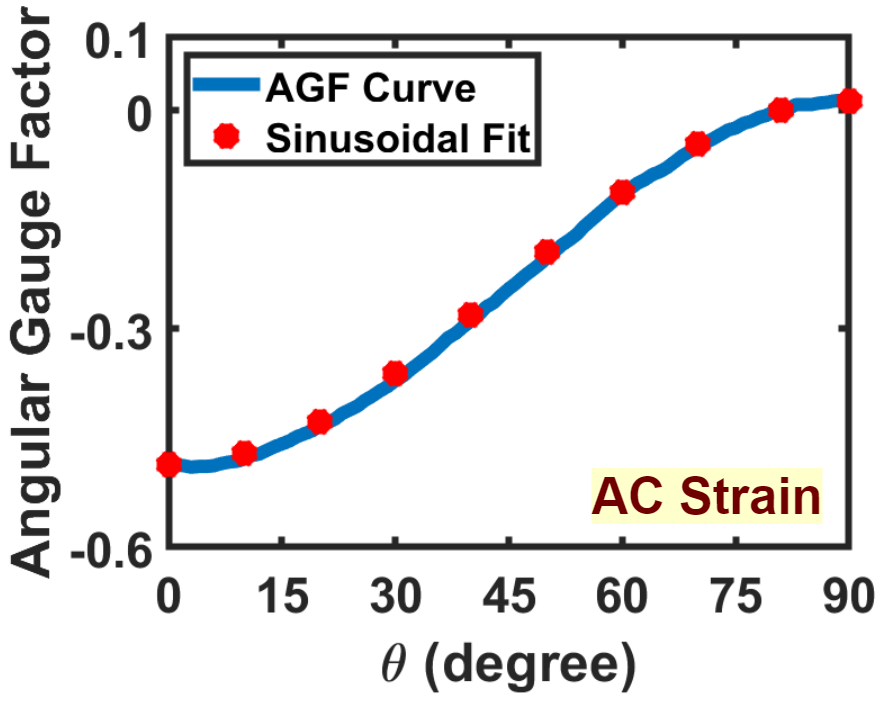}\label{P01_9a}}
\quad
    \subfigure[]{\includegraphics[height=0.18\textwidth,width=0.225\textwidth]{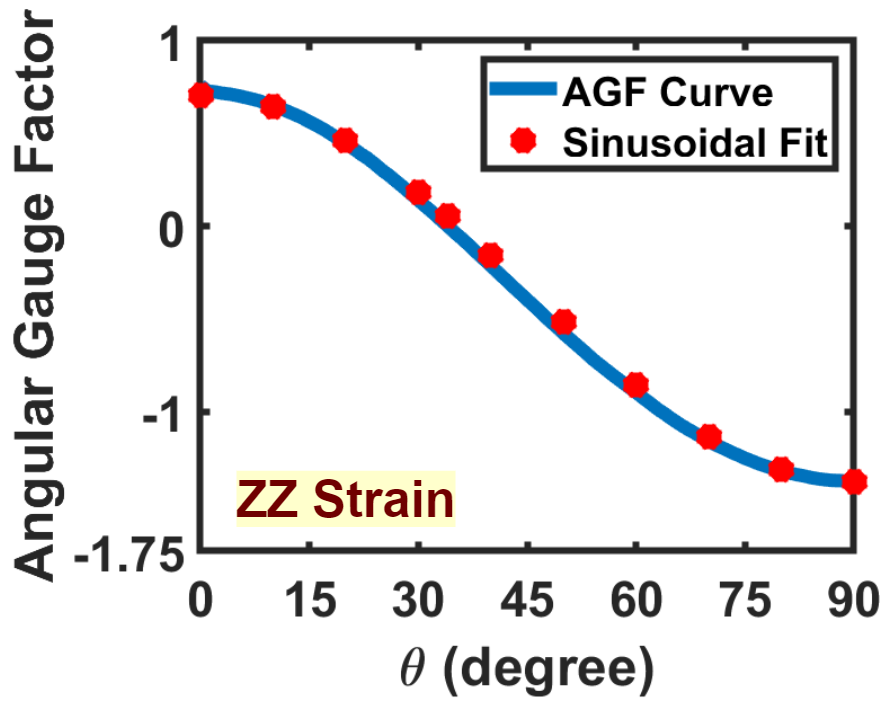}\label{P01_9b}}
\quad
    \subfigure[]{\includegraphics[height=0.18\textwidth,width=0.228\textwidth]{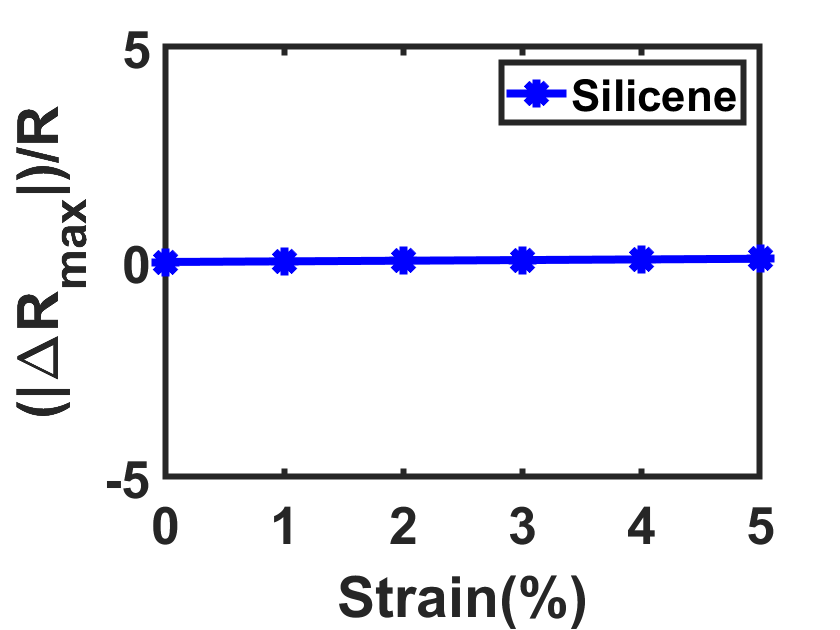}\label{P01_9c}}
\quad
    \subfigure[]{\includegraphics[height=0.18\textwidth,width=0.228\textwidth]{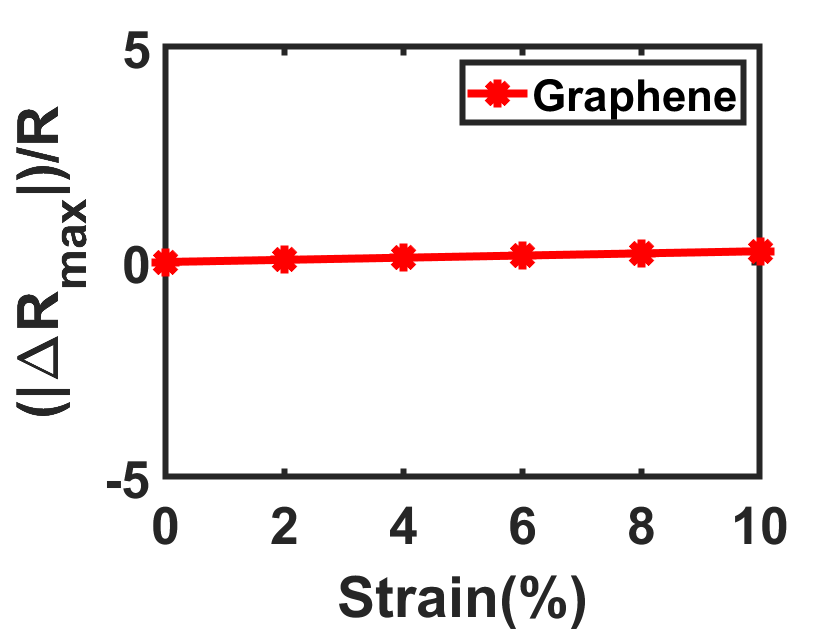}\label{P01_9d}}
 \caption{Plots of angular gauge factor (AGF) along with its sinusoidal fit as a function of transport angle $(\theta)$ along the (a) armchair, and (b) zigzag directions.  Variation of the normalized resistance with respect to strain for (a) silicene, and (b) graphene.}
\label{P01_9}
\end{figure}
\indent The maximum change in normalized resistance of silicene is plotted as a function of the percentage strain in Fig.~\ref{P01_9c}. Similarly, the plot for the maximum change in normalized resistance of ballistic graphene is shown in Fig.~\ref{P01_9d}, from the value of GF obtained by Sinha~\textit{et al.}~\cite{sinha2019piezoresistance}. From the plots, we infer that silicene is strongly resistant to a change in resistance due to strain thereby demonstrating a robust electronic characteristic. Furthermore, silicene, which has high conductivity~\cite{zhang2014thermal} and high elastic limit~\cite{peng2013mechanical}, can be used as electrodes and interconnects in flexible electronic devices. Due to the industrial compatibility of silicene with silicon fabrication technology~\cite{zhao2016rise}, silicene has the potential to beat graphene as an ideal choice for flexible electronics.\\
\section{Conclusion} \label{section_4}
In this paper, we investigated the piezoresistance of silicene using \textit{ab-initio} simulation and quantum transport theory. We calculated the directional piezoresistance for strain along the armchair and zigzag directions. We obtained a typically small value of the directional piezoresistance (magnitude less than 1), which depended sinusoidally on the transport angle. The strain insensitive transport angles corresponding to the zero gauge factors are $81^{\circ}$ and $34^{\circ}$ for armchair and zigzag strains, respectively. The small gauge factor of silicene was attributed to its robust Dirac cone and strain-independent valley degeneracy.
Based on the obtained results, we proposed silicene as an interconnect in flexible electronics. Further, this work can serve as a template for exploring flexible electronics devices and their applications using other 2D-Xenes.
\begin{acknowledgements}
The Research and Development work undertaken in the project under the Visvesvaraya Ph.D. Scheme of Ministry of Electronics and Information Technology, Government of India, is implemented by Digital India Corporation (formerly Media Lab Asia). This work was also supported by the Science and Engineering Research Board (SERB), Government of India, Grant No. CRG/2021/003102 and Grant No. STR/2019/000030.
\end{acknowledgements}
\bibliography{reference}
\end{document}